\documentclass[letterpaper, 10 pt, conference]{ieeeconf}
\usepackage{graphicx} 

\IEEEoverridecommandlockouts                              %
\let\labelindent\relax

\usepackage[utf8]{inputenc}
\usepackage{graphics} 
\usepackage{amsmath} 
\usepackage{amsthm}
\usepackage{amssymb}  
\usepackage{amsfonts}
\usepackage{comment}
\usepackage{enumitem}
\usepackage{cases,xspace,enumitem}
\usepackage{algorithm}
\usepackage{algpseudocode}

\usepackage{version}
\usepackage{stmaryrd}

\includeversion{arxiv}\excludeversion{asc}

\newtheorem{theorem}{Theorem}
\newtheorem{corollary}[theorem]{Corollary}
\newtheorem{lemma}[theorem]{Lemma}
\newtheorem{remark}[theorem]{Remark}

\newtheorem{assumption}{Assumption}

\usepackage{hyperref}%
\hypersetup{colorlinks=true, linkcolor=blue, breaklinks=true, urlcolor=blue}

\newcommand{\until}[1]{\{1,\dots, #1\}}
\newcommand{\subscr}[2]{#1_{\textup{#2}}}

\newcommand{\setdef}[2]{\{#1 \; | \; #2\}}

\newcommand{\map}[3]{#1: #2 \rightarrow #3}

\newcommand{\real}{\ensuremath{\mathbb{R}}}

\newcommand\oprocendsymbol{\hbox{$\triangle$}}
\newcommand\oprocend{\relax\ifmmode\else\unskip\hfill\fi\oprocendsymbol}

\renewcommand{\theenumi}{(\roman{enumi})}

\DeclareSymbolFont{bbold}{U}{bbold}{m}{n}
\DeclareSymbolFontAlphabet{\mathbbold}{bbold}

\usepackage[prependcaption,colorinlistoftodos]{todonotes}

\newcommand*{\mydoi}[1]{\href{http://dx.doi.org/#1}{\includegraphics[width=.75em]{doi.png}}}

\begin{asc}
\end{asc}

\begin{arxiv}
\end{arxiv}

\newcommand{\jac}[1]{D\mkern-0.75mu{#1}}

\newcommand{\seminorm}[1]{{\left\vert\kern-0.25ex\left\vert\kern-0.25ex\left\vert #1
		\right\vert\kern-0.25ex\right\vert\kern-0.25ex\right\vert}}

\newcommand{\semimeasure}[1]{\mu_{\seminorm{\cdot}}\kern-0.5ex\left(#1\right)}

\newcommand{\mcX}{\mathcal{X}}

\newcommand{\mcA}{\mathcal{A}}

\newcommand{\matrixlemma}{\ref{lemma:interval-facts}\ref{item:matmul}\xspace}

\newcommand{\mzr}[1]{\subscr{\lceil#1\rceil}{Mzr}}

\newcommand{\relu}{\mathrm{ReLU}}

\newcommand{\xover}{\overline{x}}
\newcommand{\xunder}{\underline{x}}

\title{Verifying Closed-Loop Contractivity of Learning-Based Controllers via Partitioning}
\author{Alexander Davydov
\thanks{Department of Mechanical Engineering and Ken Kennedy Institute, Rice University, Houston, TX, USA. {\tt\footnotesize davydov@rice.edu}}
}

\begin{document}

\maketitle

\begin{abstract}
    We address the problem of verifying closed-loop contraction in nonlinear control systems whose controller and contraction metric are both parameterized by neural networks. By leveraging interval analysis and interval bound propagation, we derive a tractable and scalable sufficient condition for closed-loop contractivity that reduces to checking that the dominant eigenvalue of a symmetric Metzler matrix is nonpositive. We combine this sufficient condition with a domain partitioning strategy to integrate this sufficient condition into training. The proposed approach is validated on an inverted pendulum system, demonstrating the ability to learn neural network controllers and contraction metrics that provably satisfy the contraction condition.
\end{abstract}

\section{Introduction}

\textbf{Problem motivation.} 
Designing and verifying controllers for nonlinear systems using first-principles tools is often computationally intractable. 
As a result of these challenges, it has become increasingly common to employ neural networks as feedback controllers and to train them to promote closed-loop stability, e.g., by penalizing instability at sampled states~\cite{DS-SJ-CF:21}. However, these empirical training techniques typically do not guarantee that the learned controller will generalize closed-loop stability properties to regions of state space away from these data samples.

To address this gap, a growing body of research has begun exploring formal verification of (i) Lyapunov stability~\cite{abate2020formal,yang2024lyapunov}, (ii) barrier functions for safety guarantees~\cite{peruffo2021automated}, and (iii) differential contraction properties~\cite{li2025neural}. While these works offer promising approaches toward theoretical guarantees, they generally draw upon computationally expensive primitives including satisfiability modulo theories (SMT) or branch and bound (BnB) algorithms, both of which are known to face significant scalability challenges in higher-dimensional or more complex systems.

In this work, we focus on learning both a feedback controller and a neural contraction metric that jointly guarantee closed-loop contraction in a prescribed region of interest. Our approach builds on scalable sufficient conditions for contraction which avoid the computational burden of using SMT or BnB-based verification. Contraction analysis is particularly appealing because establishing contraction yields strong corollaries including input-to-state stability, robustness to unmodeled dynamics, and robustness to stochastic disturbances~\cite{HT-SJC-JJES:21,AD-FB:24g}. 

\textbf{Related work.} One related line of research is focused on learning contracting dynamical systems from data~\cite{mohammadi2023neural,SJ-AD-DL-AKS-FB:24c}. In these works, the task is to learn a dynamical system $\dot{x} = f(x)$ which is guaranteed to be contracting everywhere in state space. In comparison, in this work, we are given a known control system $\dot{x} = f(x) + Bu$ and instead are trying to learn a feedback controller and contraction metric that ensures contraction in a subset of the state space.

The work that is most closely related to ours is~\cite{MZ-LX-GFT:24} where the authors use the theory of control contraction metrics~\cite{IRM-JJES:17} to ensure closed-loop contraction with a neural network-based controller and neural contraction metric. The authors use a Gershgorin-type sufficient condition in Theorem~2 to provide a finite-dimensional scalar condition as a sufficient condition for contraction. Moreover, in Proposition~1, they parametrize the neural contraction metric to enforce certain Killing vector field conditions which appear in the strong control contraction metric conditions~\cite[Section~III.A]{IRM-JJES:17}.

\textbf{Contributions.} We develop a scalable approach for verifying contraction of nonlinear control systems with neural network controllers and neural contraction metrics. First, we introduce a Metzler majorant sufficient condition that reduces the infinite-dimensional contraction inequality to checking the dominant eigenvalue of a single symmetric Metzler matrix. This sharpens the Gershgorin-type bound of~\cite{MZ-LX-GFT:24} and circumvents expensive SMT or BnB tools.

Second, we combine interval bound propagation, interval arithmetic, and interval Jacobian bounds to construct overapproximations of all terms in the contraction inequality. 

We additionally propose a domain-partitioning strategy that reduces conservatism and yields verifiable certificates on large regions of the state space. Finally, we integrate our verification procedure into training, enabling direct learning of closed-loop contracting controllers. A numerical experiment on the control of an inverted pendulum illustrates the effectiveness of the proposed approach.

\begin{asc}
Due to space limitations, we omit proofs and present them in our technical report.\footnote{\url{https://davydovalexander.github.io/files/ACSSC25.pdf}}
\end{asc}

\begin{arxiv}
    In this extended technical report, we present proofs of all results and include more remarks throughout.
\end{arxiv}

\section{Preliminaries}
We collect the necessary background on contraction theory and interval methods used throughout the paper.

\textbf{Contraction theory.}
We consider the dynamical system 
\begin{equation}\label{eq:nonlinear-dynamics}
    \dot{x} = f(x)
\end{equation} where $x(t) \in \real^n$ for all $t$ and $\map{f}{\real^n}{\real^n}$ is continuously differentiable. Contraction aims to address the question of when all solution trajectories of~\eqref{eq:nonlinear-dynamics} converge to one another exponentially quickly in some metric. For arbitrary metrics, this question is intractable. As a result, it is common to restrict oneself to metrics induced by Riemannian metrics on $\real^n$. To be specific, a matrix-valued mapping $\map{M}{\real^n}{\real^{n \times n}}$ satisfying $M(x) = M(x)^\top \succ 0$ for all $x$ induces a Riemannian metric on $\real^n$ and the distance between two points $x,y \in \real^n$ is defined by the length of the geodesic connecting $x$ and $y$, see~\cite{simpson2014contraction} for more details.

One of the key results in contraction theory is that contraction with respect to the Riemannian metric can be established via a matrix inequality in terms of $M, \jac{f}, $ and $\dot{M}_f$, which we highlight below.

\begin{lemma}[{Contraction inequality~\cite{WL-JJES:98}}] \label{def:contraction}
The dynamical system~\eqref{eq:nonlinear-dynamics} is \emph{contracting} with rate $c > 0$ if there exists a continuously-differentiable matrix-valued map $\map{M}{\real^n}{\real^{n \times n}}$ and two constants $a_0, a_1>0$ such that for all $x \in \real^n$, $M(x) = M(x)^\top$ and $a_0 I_n \preceq M(x) \preceq a_1 I_n$ and additionally satisfies for all $x$
\begin{equation}\label{eq:contraction-condition}
    M(x) \jac{f}(x) + \jac{f}(x)^\top M(x) + \dot{M}_f(x) \preceq -2cM(x).
\end{equation}
\end{lemma}

$\dot{M}_f$ denotes the Lie derivative of $M$ along the vector field $f$ and it is defined componentwise by $[\dot{M}_f(x)]_{ij} = \nabla M_{ij}(x)^\top f(x)$.
In essence, Lemma~\ref{def:contraction} ensures that for any pair of initial conditions, $x_0,y_0 \in \real^n$, the solutions $x(t), y(t)$ of~\eqref{eq:nonlinear-dynamics} initialized at these initial conditions satisfy $d(x(t),y(t)) \leq e^{-ct}d(x_0,y_0),$ where $d$ is the geodesic distance induced by $M$.

\textbf{Interval arithmetic.} For two vectors $\xunder,\xover \in \real^n$, we will use the partial order $\xunder \leq \xover$ if and only if $\xunder_i \leq \xover_i$ for all $i \in \until{n}$. We use an analogous partial order on the set of matrices $\real^{n \times m}$ (not to be confused with the partial order $\preceq$ on the set of symmetric matrices). For two vectors (or matrices) $\xunder \leq \xover$, we define the hyperrectangle $[\xunder,\xover] := \setdef{x \in \real^n}{\xunder \leq x\leq \xover}$. We recall the following useful facts from interval arithmetic:

\begin{lemma}\label{lemma:interval-facts}
    Suppose $a \in [\underline{a},\overline{a}] \subseteq \real, b \in [\underline{b},\overline{b}] \subseteq \real, A \in [\underline{A},\overline{A}] \subseteq \real^{n \times k}$, and $B \in [\underline{B},\overline{B}] \subseteq \real^{k \times m}$. Then
    \begin{enumerate}
        \item $a + b \in [\underline{a} + \underline{b}, \overline{a} + \overline{b}];$
        \item $ab \in [\min\{\underline{a}\underline{b},\underline{a}\overline{b},\overline{a}\underline{b},\overline{a}\overline{b}\}, \max\{\underline{a}\underline{b},\underline{a}\overline{b},\overline{a}\underline{b},\overline{a}\overline{b}\}]$; 
        \item\label{item:matmul} $[AB]_{ij} \in \big[\sum_{p=1}^k \min\{\underline{a}_{ip}\underline{b}_{pj},\underline{a}_{ip}\overline{b}_{pj},\overline{a}_{ip}\underline{b}_{pj},\overline{a}_{ip}\overline{b}_{pj}\}, \\ \qquad\qquad\qquad \sum_{p=1}^k \max\{\underline{a}_{ip}\underline{b}_{pj},\underline{a}_{ip}\overline{b}_{pj},\overline{a}_{ip}\underline{b}_{pj},\overline{a}_{ip}\overline{b}_{pj}\}\big]$.
    \end{enumerate}
\end{lemma}
We will frequently use the result in Lemma~\matrixlemma. For succinctness, we will write $\mathrm{IMM}([\underline{A},\overline{A}], [\underline{B},\overline{B}])$ to denote the mappings which outputs the intervals from Lemma~\matrixlemma.

We also define the \emph{Metzler majorant} of a matrix $A \in \real^{n \times n}$ by $[\mzr{A}]_{ij} = |a_{ij}|$ if $i \neq j$ and $[\mzr{A}]_{ii} = a_{ii}$. The name Metzler majorant alludes to the fact that $\mzr{A}$ is Metzler even if $A$ is not (off-diagonal elements are nonnegative) and that $A \leq \mzr{A}$.

\textbf{Interval bound propagation.} To construct overapproximations of neural network outputs, we will leverage interval bound propagation (IBP)~\cite{SG-KD-etal:18} which provides constant lower and upper bounds on the intermediate and output layers of a feedforward neural network. To be precise, given a neural network $\map{N}{\real^n}{\real^m}$ parametrized by
\begin{subequations}
\begin{align}\label{eq:FFNN}
    N(x) &= \xi_{L+1}  \\
    \xi_{k+1} &= W_kz_k + b_k, \quad z_0 = x, \label{eq:xi-def} \\
    z_{k+1} &= \sigma(\xi_{k+1})
\end{align}
\end{subequations}
where $W_i$ are weight matrices, $b_i$ are bias vectors, the $\xi_i$ are pre-activation variables, and $z_i$ are post-activation variables.

Given a set of possible inputs in a hyperrectangle $\mcX = [\underline{x}, \overline{x}]$, IBP generates a hyperrectangular overapproximation of the set $N(\mcX) := \setdef{N(x)}{x \in \mcX}$. To be specific, letting $[W]^+ = \relu(W)$ and $[W]^{-} = W - [W]^+$ and assuming that $\sigma$ acts entrywise and is monotone increasing, IBP provides the outputs
\begin{subequations}
\begin{align}
    \underline{\xi}_{k+1} &= [W_k]^+ \underline{z}_k + [W_k]^- \overline{z}_k + b_k \label{eq:lower-xi}\\
    \overline{\xi}_{k+1} &= [W_k]^+ \overline{z}_k + [W_k]^- \underline{z}_k + b_k \label{eq:upper-xi}\\
    \underline{z}_{k+1} &= \sigma(\underline{\xi}_{k+1}) \\
    \overline{z}_{k+1} &= \sigma(\overline{\xi}_{k+1})
\end{align}
\end{subequations}
with the initial bounds $\underline{z}_0 = \underline{x}$ and $\overline{z}_0 = \overline{x}$. Then IBP provides the estimate $N(\mcX) \subseteq [\underline{\xi}_{L+1}, \overline{\xi}_{L+1}]$. Notably, when $\underline{x} = \overline{x} = x$, then $\underline{\xi}_{L+1} = \overline{\xi}_{L+1} = \xi_{L+1}$, so we can see that as the set of possible inputs shrinks, the IBP overapproximation of the output set shrinks as well.

\section{Technical Results}

\textbf{Verification of infinite-dimensional matrix inequalities.} The main challenge in verification of contraction of a nonlinear system is that the condition~\eqref{eq:contraction-condition} must be verified at infinitely many $x$. Namely, the condition~\eqref{eq:contraction-condition} is an infinite-dimensional linear matrix inequality in $M$. To abstract away the details of the specific form of the contraction inequality~\eqref{eq:contraction-condition}, we will investigate arbitrary matrix inequalities of the form $\mcA(x) \preceq 0$ where $\map{\mcA}{\mcX}{\real^{n \times n}}$ is a matrix-valued mapping satisfying $\mcA(x) = \mcA(x)^\top$ for all $x \in \mcX$. Our main theoretical result is the following one.

\begin{theorem}\label{thm:main-result}
    Consider a matrix-valued mapping $\map{\mcA}{\mcX}{\real^{n \times n}}$ satisfying $\mcA(x) = \mcA(x)^\top$ for all $x$. Suppose there exists a symmetric $G \in \real^{n \times n}$ such that $G \geq \mzr{\mcA(x)}$ for all $x \in \mcX$. If $\lambda_{\max}(G) \leq 0$, then $\mcA(x) \preceq 0$ for all $x \in \mcX$.
\end{theorem}

\begin{arxiv}
\begin{proof}
Let $v \in \real^n$ and let $x \in \mcX$. Then
\begin{align*}
    v^\top \mcA(x)v &= \sum_{i=1}^n v_i^2 \mcA(x)_{ii} + \sum_{i=1}^n \sum_{j=1, j\neq i}^n v_iv_j \mathcal{A}(x)_{ij} \\
    &\leq \sum_{i=1}^n v_i^2 \mcA(x)_{ii} + \sum_{i=1}^n \sum_{j=1, j\neq i}^n |v_i| |v_j| |\mathcal{A}(x)_{ij}| \\
    &\leq \sum_{i=1}^n v_i^2 G_{ii} + \sum_{i=1}^n \sum_{j=1, j\neq i}^n |v_i| |v_j| G_{ij} \\
    &= |v|^\top G |v| \leq 0,
\end{align*}
where we denote by $|v|$ the entrywise absolute value of $v$. We have used that $G \geq \mzr{\mcA(x)}$ and that $\lambda_{\max}(G) \leq 0$. 
\end{proof}
\end{arxiv}

Theorem~\ref{thm:main-result} demonstrates that a sufficient condition for the infinite-dimensional matrix inequality $\mcA(x) \preceq 0$ can be verified by finding the dominant eigenvalue of a constant Metzler matrix, which is much more tractable. Of course, there are even constant matrices $\mcA$ which are negative semidefinite but whose Metzler majorant is not. Therefore, the sufficient condition in Theorem~\ref{thm:main-result} trades off computational efficiency for tightness. 

\textbf{Interval bounds on Jacobians of neural networks.} Consider a feedforward neural network $\map{N}{\real^n}{\real^m}$ given by~\eqref{eq:FFNN}. Since the contraction condition~\eqref{eq:contraction-condition} involves Jacobians of mappings, it will be useful to have bounds on $\jac{N}(x)$ for $x \in \mcX$. First note that because of the feedforward structure of $N$, $\jac{N}(x)$ admits the form
\begin{equation}\label{eq:jacNN}
    \jac{N}(x) = W_L J_{L-1}(x) W_{L-1} J_{L-2}(x) \dots J_1(x)W_1
\end{equation}
where $J_i(x) = \jac{\sigma}(\xi_{i})$ where $\jac{\sigma}(x)$ is the diagonal matrix with $j$-th diagonal entry equal to $\sigma'(x_j)$ and $\xi_i$ is given in~\eqref{eq:xi-def}. Notably, to attain a bound on $\jac{N}(x)$, we simply need to bound the product of constant matrices (the $W_i$) and diagonal matrices (the $J_i(x)$).

To find upper and lower bounds on each $J_i(x)$, we will make the following simplifying assumption:

\begin{assumption}\label{assm:mono-derivative}
    Suppose $\sigma$ is twice continuously differentiable and $\sigma''(x) \geq 0$ for all $x$.
\end{assumption}
In addition to monotonicity of $\sigma$, Assumption~\ref{assm:mono-derivative} imposes convexity on $\sigma$ as well. Common activation functions which satisfy this assumption include the softplus activation, $\sigma(x) = \log(1 + \exp(x))$ and the smooth leaky ReLU, $\sigma(x) = \alpha x + (1-\alpha)\log(1+\exp(x))$ with $\alpha \in (0,1)$. Under Assumption~\ref{assm:mono-derivative}, we get the following bounds on each $J_i(x)$:
\begin{lemma}\label{lemma:diag-bounds}
    Suppose Assumption~\ref{assm:mono-derivative} holds. Then $\underline{J}_i \leq J_i(x) \leq \overline{J}_i$ where
    \begin{equation}
        \underline{J}_i = \jac{\sigma}(\underline{\xi}_i), \qquad \overline{J}_i = \jac{\sigma}(\overline{\xi}_i)
    \end{equation}
    and $\underline{\xi}_i, \overline{\xi}_i$ are given in the IBP algorithm~\eqref{eq:lower-xi},~\eqref{eq:upper-xi}.
\end{lemma}

Lemma~\ref{lemma:diag-bounds} highlights how the IBP bounds for pre-activation variables immediately provide us with bounds on the $J_i(x)$. These bounds improve on those in~\cite{MZ-LX-GFT:24} where they instead use $\underline{J}_i = aI, \overline{J}_i = bI$ where $0<a \leq \sigma'(x) \leq b$ for all $x$.

With these bounds in hand and the fact that the $W_i$ are constant, we can recursively apply Lemma~\matrixlemma to get interval bounds on the Jacobian of the neural network. The full algorithm is presented in Algorithm~\ref{alg:jac-nn}.

\begin{algorithm}
\caption{Output bounds on $\jac{N}(x)$ for $x \in [\underline{x}, \overline{x}]$}\label{alg:jac-nn}
\begin{algorithmic}[1] 
    \State Require $x \in [\underline{x}, \overline{x}]$, $N$ feedforward neural network 
    \State Call $\mathrm{IBP}(N; \underline{x},\overline{x})$ to get bounds $\xi_i \in [\underline{\xi}_i, \overline{\xi}_i]$ for $x \in \mcX$ and $i \in \until{L+1}$
    \State \textcolor{green!50!black}{\textit{\# Create running variable that tracks matrix bounds as we multiply right to left}}
    \State Initialize matrix interval $[\underline{P}, \overline{P}]$ where $\underline{P} = \overline{P} = I_n$
    \For{$i = 1$ to $L-1$}
        \State \textcolor{green!50!black}{\textit{\# Multiply on the left by constant matrix $W_i$}}
        \State $[\underline{P}, \overline{P}] \gets \mathrm{IMM}([W_i, W_i], [\underline{P}, \overline{P}])$
        \State \textcolor{green!50!black}{\textit{\# Get diagonal bounds $\underline{J}_i, \overline{J}_i$}}
        \State $\underline{J}_i \gets \jac{\sigma}(\underline{\xi}_i), \overline{J}_i \gets \jac{\sigma}(\overline{\xi}_i)$
        \State \textcolor{green!50!black}{\textit{\# Multiply on the left by interval diagonal matrix $J_i$}}
        \State $[\underline{P},\overline{P}] \gets \mathrm{IMM}([\underline{J}_i, \overline{J}_i], [\underline{P}, \overline{P}])$
    \EndFor
    \State \textcolor{green!50!black}{\textit{\# Multiply on the left by output matrix $W_L$}}
    \State $[\underline{P}, \overline{P}] \gets \mathrm{IMM}([W_L, W_L], [\underline{P}, \overline{P}])$
    \State \Return $[\underline{P}, \overline{P}]$; optionally \Return $\{[\underline{\xi}_i, \overline{\xi}_i]\}_{i=1}^{L+1}$
\end{algorithmic}
\end{algorithm}

\begin{arxiv}
\begin{remark}\label{rmk:nonmonotone}
    In the case that Assumption~\ref{assm:mono-derivative} does not hold, then Lemma~\ref{lemma:diag-bounds} and Algorithm~\ref{alg:jac-nn} can be modified as follows. Denote by $\underline{\xi}_{i,j}, \overline{\xi}_{i,j}$ the $j$-th components of $\underline{\xi}_i$ and $\overline{\xi}_i$, respectively. Then we can define
    \begin{equation}
        [\underline{J}_i]_{jj} = \min_{y \in [\underline{\xi}_{i,j}, \overline{\xi}_{i,j}]} \sigma'(y), \qquad [\overline{J}_i]_{jj} = \max_{y \in [\underline{\xi}_{i,j}, \overline{\xi}_{i,j}]} \sigma'(y),
    \end{equation}
    and zero on the off-diagonal entries. It is easy to see that these are lower and upper bounds on each $J_i(x)$. Under Assumption~\ref{assm:mono-derivative}, these minimization and maximization problems have explicit closed-form expressions which are presented in Lemma~\ref{lemma:diag-bounds}.
\end{remark}

Suppose we want to provide bounds on $B\jac{N}(x)$ given that $N$ is a feedforward neural network, $B$ is a matrix and $x \in [\underline{x},\overline{x}]$. Then we could get the bounds by calling Algorithm~\ref{alg:jac-nn} to get bounds on $\jac{N}(x)$ and then use Lemma~\matrixlemma to get bounds on the product between $B$ and $\jac{N}(x)$. Alternatively, we could define a new feedforward neural network $\tilde{N}$ which shares all the same weights and biases except for output weight $BW_{N-1}$ and output bias $Bb_{N-1}$. It is clear that $\tilde{N}(x) = BN(x)$ and that $\jac{\tilde{N}}(x) = B\jac{N}(x)$. This means that we can run Algorithm~\ref{alg:jac-nn} on the neural network $\tilde{N}$ to get bounds on $B\jac{N}(x)$. In the IBP literature this approach is called \emph{eliding the last layer} and it has been shown to produce tighter estimates than using Lemma~\matrixlemma~\cite{SG-KD-etal:18}. Note that this procedure is only possible if the last layer is affine.
\end{arxiv}

\section{Contraction Verification}
Consider the control system 
\begin{equation}\label{eq:system}
    \dot{x} = f(x) + Bu,
\end{equation}
where $x \in \real^n$, $u \in \real^m$, $B \in \real^{n \times m}$. Suppose we train a feedforward neural network feedback controller $\map{u}{\real^n}{\real^m}$ and a \emph{neural contraction metric} $\map{M}{\real^n}{\real^{n \times n}}$ for~\eqref{eq:system}. We aim to verify whether the contraction condition~\eqref{eq:contraction-condition} is satisfied for the closed-loop dynamics $\subscr{f}{CL}(x) = f(x) + Bu(x)$ in a hyperrectangular region of interest in the state space, $\mcX := [\xunder,\xover]$. 

Define the matrix-valued mapping $\map{\mcA}{\real^n}{\real^{n\times n}}$ by 
$\mcA(x) = M(x)\jac{\subscr{f}{CL}}(x) + \jac{\subscr{f}{CL}}(x)^\top M(x) + \dot{M}_{f_{CL}}(x) + 2cM(x).$
In view of Definition~\ref{def:contraction}, contraction with rate $c > 0$ with respect to metric induced by $M$ is guaranteed provided $\mcA(x) \preceq 0$. For simplicity, going forward, we will take $c = 0$, but the following analysis will work just as well with $c > 0$.

We can decompose $\mcA$ into the component coming from the dynamics and the one from the controller, i.e.,
\begin{align*}
    \mcA(x) &= \mcA_{f}(x) + \mcA_u(x), \qquad \qquad \text{where} \\
    \mcA_f(x) &= M(x)\jac{f}(x) + \jac{f}(x)^\top M(x) + \dot{M}_f(x) \\
    \mcA_u(x) &= M(x)B\jac{u}(x) + \jac{u}^\top B^\top M(x) + \dot{M}_{Bu}(x)
\end{align*}

In the end, our goal will be to construct a constant matrix $G$ for which $G \geq \mzr{\mcA(x)}$ for all $x \in \mcX$. To this end, we make the following standing assumption on the dynamics
\begin{assumption}
    For all $\underline{x} \leq \overline{x}$, there exist $\underline{f}(\underline{x},\overline{x}), \overline{f}(\underline{x},\overline{x}) \in \real^{n}$ and $\underline{\jac{f}}(\underline{x},\overline{x}), \overline{\jac{f}}(\underline{x},\overline{x}) \in \real^{n\times n}$ such that for all $x \in \mcX$
    \begin{align}
        \underline{f}(\underline{x},\overline{x}) \leq \;&f(x) \leq \overline{f}(\underline{x},\overline{x}) \\
        \underline{\jac{f}}(\underline{x},\overline{x}) \leq \;&\jac{f}(x) \leq \overline{\jac{f}}(\underline{x},\overline{x}).
    \end{align}
\end{assumption}
For simple dynamical systems, these bounds may be analytically computable in terms of $\underline{x},\overline{x}$, e.g., as we will do in Section~\ref{sec:experiments}. Otherwise, for some more complex dynamical systems, dedicated software for interval arithmetic may be used, e.g., \texttt{npinterval}~\cite{harapanahalli2023toolbox}. When the bounds $\underline{x}$ and $\overline{x}$ are clear from context, we will simply write $\underline{f}, \overline{f}, \underline{\jac{f}}, \overline{\jac{f}}$.

Since $u$ is a feedforward neural network, we can directly apply IBP and Algorithm~\ref{alg:jac-nn} to attain bounds on $u(x)$ and $\jac{u}(x)$ for $x \in [\underline{x}, \overline{x}]$. 
On the other hand, since $M$ must be constrained to be positive definite for all inputs, we need to tailor IBP and its variation to effectively bound $M(x)$ and $\dot{M}_f(x),$ and $\dot{M}_{Bu}(x)$ in terms of the bounds $x \in [\underline{x}, \overline{x}]$.

\textbf{Neural contraction metric bounds.}
To enforce positive definiteness of $M(x)$ for all $x$, we use the same parametrization for the neural contraction metric as used in~\cite{DS-SJ-CF:21}. Namely, the contraction metric is parametrized via 
\begin{equation}\label{eq:NCM}
    M(x) = N(x)^\top N(x) + \varepsilon I_n,
\end{equation}
where $\varepsilon > 0$ is a small constant enforcing strict positive definiteness of $M$ and $\map{N}{\real^n}{\real^{n \times n}}$ is parametrized by a feedforward neural network where the output has been reshaped to be an $n \times n$ matrix.

In addition to bounds on $M(x)$ for $x \in \mcX$, for the computation of the terms of $\dot{M}_f$ and $\dot{M}_{Bu}$, we need estimates on each of the $\nabla M_{ij}(x)$ for $i,j \in \until{n}$. Since $M_{ij}(x) = \varepsilon\delta_{ij} + \sum_{k=1}^n N_{ki}(x)N_{kj}(x)$, where $\delta_{ij}$ is the Kronecker delta, we can compute
\begin{align}
    \nabla  M_{ij}(x) 
    = \jac{N_i}(x)^\top N_j(x) + \jac{N_j}(x)^\top N_i(x),\label{eq:vector-gradient}
\end{align}
where $N_i(x)$ denotes the $i$-th column of the matrix $N(x)$ and $\jac{N}_i(x)$ is the Jacobian of the map $N_i$ evaluated at $x$. 

To get bounds on $M(x)$ and each of the $\nabla M_{ij}(x)$, we apply IBP to the network $N$, reshape the last-layer bounds to get matrix-valued bounds and then use the interval Jacobian procedure in Algorithm~\ref{alg:jac-nn} to each column, $N_i(x)$. Combining these bounds using~\eqref{eq:vector-gradient} and Lemma~\matrixlemma provides interval bounds on $\nabla M_{ij}(x)$.

\begin{arxiv}
\begin{remark}
    In certain cases, it may be sufficient to use a constant $M$ instead of a state-varying one. In this case, bounds on $M$ and $\nabla M_{ij}$ are immediate since $M$ is constant and $\nabla M_{ij} = 0$ for all $i,j \in \until{n}$.
\end{remark}
\end{arxiv}

\newcommand{\mdfupper}{U_f}
\newcommand{\mdflower}{L_f}
\newcommand{\mbduupper}{U_u}
\newcommand{\mbdulower}{L_u}

\textbf{Putting all the bounds together.}
We are now prepared to combine all the bounds together to construct the matrix $G$.

First, given the bounds on $\nabla M_{ij}(x)$, we can get bounds on $\dot{M}_f(x)$ and $\dot{M}_{Bu}(x)$ by applying Lemma~\matrixlemma to the inner products between the interval matrices $[\underline{\nabla M}_{ij}, \overline{\nabla M}_{ij}]$ and either $[\underline{f}, \overline{f}]$ or $[\underline{Bu}, \overline{Bu}]$. These products give us bounds $\dot{M}_f(x) \in [\underline{\dot{M}_f}, \overline{\dot{M}}_f]$ and $\dot{M}_{Bu}(x) \in [\underline{\dot{M}}_{Bu}, \overline{\dot{M}}_{Bu}]$. 

Finally by two more applications of Lemma~\matrixlemma, we can construct bounds $M(x)\jac{f}(x) \in [\mdflower, \mdfupper]$ and $M(x)B\jac{u}(x) \in [\mbdulower, \mbduupper]$. Recalling the definitions then of $\mcA_f$ and $\mcA_u$, we attain the bounds
\begin{align*}
    \mcA_f(x) &\in [\mdflower + \mdflower^\top + \underline{\dot{M}_f}, \mdfupper + \mdfupper^\top + \overline{\dot{M}_f}] =: [\underline{\mcA_f}, \overline{\mcA}_f] 
    \\
    \mcA_u(x) &\in [\mbdulower + \mbdulower^\top + \underline{\dot{M}}_{Bu}, U_u + U_u^\top + \overline{\dot{M}}_{Bu}] =: [\underline{\mcA}_u, \overline{\mcA}_u]
\end{align*}

Finally, $\mcA(x) \in [\underline{\mcA}_f + \underline{\mcA}_u, \overline{\mcA}_f + \overline{\mcA}_u] =: [\underline{\mcA}, \overline{\mcA}]$. 

To apply Theorem~\ref{thm:main-result}, we need to construct a matrix $G$ such that $\mzr{\mcA(x)} \leq G$. We can do this using the bounds on $\mcA(x)$. We define the matrix one element at a time:
\begin{equation}\label{eq:G-def}
    G_{ij} = \begin{cases}
        \max\{\overline{\mcA}_{ij}, -\underline{\mcA}_{ij}\}, \quad& \text{ if } i \neq j \\
        \overline{\mcA}_{ij}, \quad & \text{ if } i = j
    \end{cases}.
\end{equation}
By construction, $\mzr{\mcA(x)} \leq G$ for all $x \in \mcX$ by definition of $\underline{\mcA}, \overline{\mcA}$. Then we arrive at our main sufficient condition for verifying contraction on a hyperrectangular domain.
\begin{corollary}\label{cor:G-to-contraction}
    If $\lambda_{\max}(G) \leq 0$, then $\mcA(x) \preceq 0$ for all $x \in \mcX$. In particular, the closed-loop dynamics are contracting with respect to the neural contraction metric $M$ in~\eqref{eq:NCM}.
\end{corollary}
\begin{arxiv}
\begin{proof}
    First we note that by construction of $M$, we have that $M(x) \succeq \varepsilon I_n$. Moreover, by continuity of $M$ and compactness of the domain $\mcX$, we also have that $M(x) \preceq RI_n$ for some $R > 0$. Moreover, since $\lambda_{\max}(G) \leq 0$ and $G \geq \mzr{\mcA(x)}$ for all $x \in \mcX$, we also have that $\mcA(x) \preceq 0$ by Theorem~\ref{thm:main-result}. As a result, we verify contraction with respect to the neural contraction metric $M$.
\end{proof}
\end{arxiv}

We summarize the full algorithm from going from input bounds $x \in [\underline{x},\overline{x}]$ to the construction of $G$ in Algorithm~\ref{alg:full-algorithm}.

\begin{algorithm}
\caption{Construction of $G$ matrix from $x \in [\underline{x},\overline{x}]$}\label{alg:full-algorithm}
\begin{algorithmic}[1] 
    \State Require $x \in [\underline{x}, \overline{x}]$, $N$ and $u$ are feedforward neural networks 
    \State Get bounds $f(x) \in [\underline{f},\overline{f}]$ and $\jac{f}(x) \in [\underline{\jac{f}},\overline{\jac{f}}]$
    \State Call Algorithm~\ref{alg:jac-nn} to get bounds $Bu(x) \in [\underline{Bu}, \overline{Bu}]$, $B\jac{u}(x) \in [\underline{B\jac{u}},\overline{B\jac{u}}]$
    \State Use Algorithm~\ref{alg:jac-nn} and Lemma~\matrixlemma to get bounds $M(x) \in [\underline{M}, \overline{M}]$, $\nabla M_{ij}(x) \in [\underline{\nabla M}_{ij},\overline{\nabla M}_{ij}]$ for $i,j \in \until{n}$
    \For{$(i,j) \in \until{n}^2$}
            \State $\big[[\underline{\dot{M}}_f]_{ij}, [\overline{\dot{M}}_{f}]_{ij}\big] \gets \mathrm{IMM}([\underline{\nabla M}_{ij}^\top,\overline{\nabla M}_{ij}^\top], [\underline{f}, \overline{f}])$
            \State $\big[[\underline{\dot{M}}_{Bu}]_{ij}, [\overline{\dot{M}}_{Bu}]_{ij}\big] \gets \mathrm{IMM}([\underline{\nabla M}_{ij}^\top,\overline{\nabla M}_{ij}^\top], [\underline{Bu}, \overline{Bu}])$
    \EndFor
    \State \textcolor{green!50!black}{\textit{\# Get bounds on $M(x)\jac{f}(x)$ and $M(x)B\jac{u}(x)$}}
    \State $[\mdflower,\mdfupper] \gets \mathrm{IMM}([\underline{M},\overline{M}], [\underline{\jac{f}},\overline{\jac{f}}])$
    \State $[\mbdulower,\mbduupper] \gets \mathrm{IMM}([\underline{M},\overline{M}], [\underline{B\jac{u}},\overline{B\jac{u{}}}])$
    \State Construct $G$ via~\eqref{eq:G-def}
    \State \Return $G$
\end{algorithmic}
\end{algorithm}

\newcommand{\under}[1]{\underline{#1}}
\newcommand{\ov}[1]{\overline{#1}}
\section{Domain Partitioning and Training Algorithm}

\textbf{Domain partitioning.}
For the time being, we have assumed that $\mcX = [\xunder,\xover]$ and that we run IBP over this domain. However, it is known that for deep networks, IBP can be conservative, especially when the input domain $\mcX$ is large. Similarly, the interval matrix multiplication bound in Lemma~\matrixlemma can also be conservative when the sets $[\underline{A},\overline{A}]$ and $[\under{B},\ov{B}]$ are large. However, as the sets become smaller, the estimates also become tighter. This observation motivates the idea of \emph{partitioning the domain of interest}.

Suppose we can partition our domain of interest $\mcX$ into the union of hyperrectangles, i.e., $\mcX = \cup_{i=1}^m [\under{x}^i, \xover^i]$
. Then if we can verify that~\eqref{eq:contraction-condition} is satisfied for each $x \in [\xunder^i,\xover^i]$ and for each $i \in \until{m}$, then the closed-loop dynamics are contracting on all of $\mcX$. Then for each $i \in \until{m}$, we can apply Algorithm~\ref{alg:full-algorithm} on the domain $[\xunder^i,\xover^i]$ to construct a constant $G^i$. Moreover, if each $G^i$ satisfies $\lambda_{\max}(G^i) \leq 0$, then we have verified contraction on the entirety of $\mcX$. Intuitively, rather than checking that the eigenvalues of a single symmetric matrix $G$ are nonpositive, we now check that a family of matrices, $G^i$ all have nonpositive eigenvalues, where each matrix $G^i$ is used to verify the contraction inequality~\eqref{eq:contraction-condition} on the subdomain $[\xunder^i,\xover^i]$.

For verification, this insight can be used to design an algorithm to adaptively partition the domain when contraction cannot be verified on the entirety of the $\mcX$. Suppose that a single application of Algorithm~\ref{alg:full-algorithm} yields a matrix $G$ with $\lambda_{\max}(G) > 0$. Then one can partition the domain $\mcX$ into $2^n$ hyperrectangular subdomains and run Algorithm~\ref{alg:full-algorithm} on each of these subdomains. If $\lambda_{\max}(G^i) \leq 0$ on any of these subdomains, then we have verified that~\eqref{eq:contraction-condition} holds on that subdomain. For any subdomain where $\lambda_{\max}(G^i) > 0$, we can again partition it into $2^n$ more subdomains. 

Intuitively, in the limit as the number of partitions goes to infinity, the sets $[\xunder^i,\xover^i]$ begin converging to singletons at which point IBP, Lemma~\matrixlemma, and the extension of IBP to Jacobians of neural networks become exact. Then $G^i$ will be arbitrarily close to $\mzr{\mcA(x)}$ and verifying $\lambda_{\max}(G^i) \leq 0$ will be equivalent to verifying $\lambda_{\max}(\mzr{\mcA(x)}) \leq 0$. 

\textbf{Training algorithm.} The sufficient condition based on Theorem~\ref{thm:main-result} and domain partitioning motivate the following training algorithm for formal guarantees of contraction on a domain of interest $\mcX$. First, partition $\mcX$ into a collection of hyperrectangles $\{[\xunder^i,\xover^i]\}_{i=1}^m$. Then initialize the weights and biases of the neural networks $u$ and $N$ for the controller and the neural contraction metric. Then define the loss function
\begin{equation}\label{eq:training-loss}
    \mathrm{Loss}(\theta) = \sum\nolimits_{i=1}^m \sum\nolimits_{j=1}^d \mathrm{ReLU}(\lambda_j(G^i)),
\end{equation}
where $\lambda_j(G^i)$ denotes the $j$-th largest eigenvalue of the symmetric matrix $G^i$ generated by Algorithm~\ref{alg:full-algorithm} and $\theta$ denotes the collection of trainable parameters in $u$ and $N$. It is important to note that $\mathrm{Loss}(\theta) = 0$ only if each $G^i$ has nonpositive eigenvalues and thus Corollary~\ref{cor:G-to-contraction} implies that the closed-loop dynamics are contracting on $\mcX$.

\section{Numerical Experiment}\label{sec:experiments}

\begin{figure}
\centering
\includegraphics[width=0.96\linewidth]{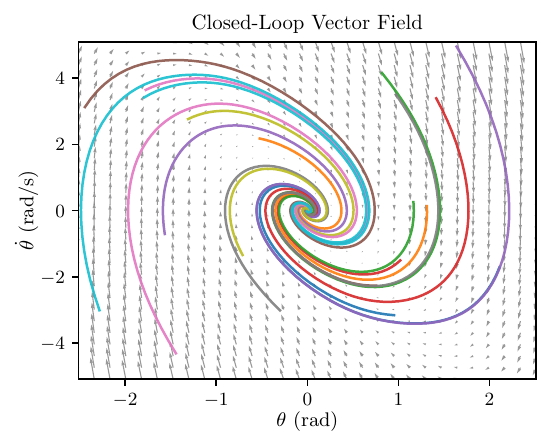}\vspace{-0.5em}
\caption{Closed-loop vector field of the inverted pendulum~\eqref{eq:inv-pend} with a neural network controller trained using the loss function~\eqref{eq:training-loss}. }\label{fig:NCM-pendulum}
\end{figure}
We consider the classical inverted pendulum
\begin{equation}\label{eq:inv-pend}
    \begin{pmatrix}
        \dot{x}_1 \\ \dot{x}_2
    \end{pmatrix} = \begin{pmatrix}
        x_2 \\ g\sin(x_1)/\ell 
    \end{pmatrix} + \begin{pmatrix}
        0 \\ 1/(m\ell^2)
    \end{pmatrix}u,
\end{equation}
where $x_1 = \theta$ denotes the angular position, $x_2 = \dot{\theta}$ denotes the angular velocity, $g$ is the gravitational constant, $m$ is the mass, and $\ell$ is the length of the pendulum. 

We train a controller with two hidden layers of width $16$ and train our NCM using~\eqref{eq:NCM} where $N$ has two hidden layers of width $32$ and $\varepsilon = 0.1$. For both networks, we use the softplus activation and for the controller network, we constrain inputs to satisfy $|u(x)| \leq 4mg\ell$ by adding a scaled $\tanh$ at the output. We train using the Adam optimizer with a learning rate of $0.01$ and we partitioned the domain into $16^2$ equal-sized hyperrectangles and used the loss function~\eqref{eq:training-loss}. Code to reproduce our results is available.\footnote{\url{https://github.com/davydovalexander/interval-bound-contraction}}

We initialize training with a small domain $\xover = (\pi/100, 0.05)$, $\xunder = -\xover$ and every time the loss hits zero, we increase the size of the domain $\xover \gets \xover + (\pi/100, 0.06)$. After 20000 epochs, we certify closed-loop contraction for $\xover = (89\pi/100, 5.33), \xunder = -\xover$. Figure~\ref{fig:NCM-pendulum} shows the closed-loop phase portrait and $20$ trajectories in phase space. 

We can pictorially see that even for this open-loop unstable system, we are able to certify contraction of the closed-loop system with a neural network controller and a neural contraction metric in a rectangular subset of the state space.

\section{Conclusions}
In this work we studied the problem of contraction verification of nonlinear systems with neural network controllers and neural contraction metrics. By leveraging interval analysis and IBP, we provide a scalable sufficient condition for contraction which corresponds to checking that all eigenvalues of a symmetric Metzler matrix are nonpositive. We integrate this sufficient condition and a domain partitioning strategy to enable learning of neural network controllers and neural contraction metrics with certificates of closed-loop contraction. We demonstrate our methodology in the control of an inverted pendulum.

Future work entails more empirical validation in larger-scale nonlinear systems. Additionally, less conservative sufficient conditions than Theorem~\ref{thm:main-result} or sharper neural network bounding methodologies than IBP could yield new training algorithms that provide sharper theoretical guarantees. Moreover, it could be of interest to verify contraction with respect to non-Euclidean norms~\cite{AD-SJ-FB:20o} or Finsler metrics~\cite{FF-RS:14} where Metzler matrices often appear in the analysis.

\begin{arxiv}
\appendices

\section{Additional Training Details}

In this section, we elaborate a bit more on some of the details for training. 

For the neural network controller, we enforce that $u(0) = 0$ over the duration of training and that $|u(x)| \leq 4mg\ell$ as well. To ensure these constraints are satisfied, we let $\tilde{N}$ be an unconstrained neural network of the form~\eqref{eq:FFNN}. Then the controller network is parametrized by $$u(x) = \mathrm{scaledtanh}(\tilde{N}(x) - \tilde{N}(0)),$$
where $\mathrm{scaledtanh}(x) = s \tanh(x/s)$, where $s$ denotes the scale factor and enforces that the output is in the interval $[-s,s]$. In the case of the inverted pendulum, since it is a single-input system, the output $u$ is scalar and we let $s = 4mg\ell$. 

We can routinely apply interval bound propagation to this parametrization of the controller because subtracting $\tilde{N}(0)$ is a constant shift at the last layer. To apply Algorithm~\ref{alg:jac-nn}, we apply the same procedure to acquire bounds on each $J_i$. The only modification is in the last layer for interval bounds on the Jacobian of $\mathrm{scaledtanh}$. We get interval bounds using Remark~\ref{rmk:nonmonotone}. In particular, the expressions in Remark~\ref{rmk:nonmonotone} can be written more concretely: for scalar $\xunder, \xover$, we have that for all $x \in [\xunder,\xover]$, $l \leq \mathrm{scaledtanh}'(x) \leq u$ where $l = \min\{\mathrm{scaledtanh}'(\xunder),\mathrm{scaledtanh}'(\xover)\}$ and $u = 1$ if $0 \in [\xunder,\xover]$ and $\max\{\mathrm{scaledtanh}'(\xunder),\mathrm{scaledtanh}'(\xover)\}$ otherwise.

In training, we start with $r^2$ equal-sized partitions (with $r=16$). We then refine the partition further by resetting $r \gets r + 2$ if the loss function does not converge to $0$ within 2000 epochs. The code is written to handle batches so we are able to compute all the eigenvalues of all the $G^i$ in parallel.
\end{arxiv}

\bibliography{alias,Main,sasha, FB}

\end{document}